# Photon-assisted-tunneling in a Coupled Double Quantum Dot out of Thermal Equilibrium

*Runan Shang[1], Haiou Li[1], Gang Cao[1], Ming Xiao[1], Tao Tu[1], Guang-Can Guo[1], Hong-Wen Jiang[2] and Guo-Ping Guo[1]*

1 Key Laboratory of Quantum Information, University of Science and Technology of China, Chinese Academy of Sciences, Hefei 230026, People's Republic of China

2 Department of Physics and Astronomy, University of California at Los Angeles, California 90095, USA.

**ABSTRACT**：We perform photon-assisted-tunneling (PAT) experiments on a GaAs double quantum dot device under high microwave excitation power. Photon-assisted absorption of up to 14 photons is observed, when electron temperature (>1K) are far above the lattice temperature. Signatures of Landau-Zener-Stückelberg (LZS) interference are found even in this non-equilibrium PAT spectrum. In addition, the charge state relaxation time $T_1 \sim 8ns$ measured in this out of thermal equilibrium double quantum dot is in agreement with other previous reports.

Electro-statically-defined semiconductor quantum dot (QD) devices serve as an excellent platform of testing the properties of tunable two-level quantum systems as charge and spin qubits[1-5]. Among a variety of useful characterization and manipulation techniques, spectroscopy of microwave excited transport, known as photon assisted tunneling (PAT), are now used routinely to study the energy spectra and dynamic effects in QDs [6-8].

In this Letter, we report the study of a GaAs double quantum dot (DQD) device under high microwave power. In the presence of such strong excitation, the electron system is not in thermal equilibrium with the crystal lattice and thus they have different temperature. Thermal non-equilibrium effect in mesoscopic system has attracted increasing attention recently since the small scale of current electronic devices makes it possible for them to be driven out of thermal equilibrium using small power. We would like to see how various interesting dynamic effects (such as energy relaxation) of the two-level system change in thermal non-equilibrium cases where the electron temperature is significantly different from the lattice temperature.

Our sample is fabricated on a $GaAs/Al_{0.3}Ga_{0.7}As$ heterostructure which contains a two-dimensional electron gas (2DEG) lying 100nm below the wafer surface with electron density $2.0 \times 10^{11} cm^{-2}$ and mobility $6 \times 10^4 \, cm^2/V \cdot s$. This sample is placed in a Helium-3 refrigerator[10] which could reach a base temperature of 240mK and hold on to it for 2-3 days until it has to be warmed up to 10K and cooled down again. Gates 1-6 shown in Fig.1 (a) form a typical DQD by depleting the electrons in 2DEG with Ti/Au top gates and gate 7 modulates the quantum point contact (QPC) channel to detect the charge state in the DQD by reading out the QPC differential current using a standard lock in measurement.

(N, M) is used to represent the state with N electrons in the left dot and M electrons in the right dot. We estimate that there are about 10 electrons in each dot by measuring the charging energy. Of course, in principle, many-body states exist in this coupled DQD. However, we can still simply describe the system as a two-level one in which the topmost level of each dot is occupied by one electron and the others stay in their respective ground states.

When the frequency of microwave photon matches the energy separation of the two-level system at zero DQD source-drain bias, electrons on the lower (upper) energy level can absorb (emit) one photon and jumped to the upper (lower) energy level. The resulting charge state repopulation could be detected by the QPC[14]. When the energy separation is n-times of the frequency of photon, the electron can absorb or emit multiple photons so the state transition would occur at large detuning. This is the so-called n-γ process.

Fig.1 (b) shows a very promising region between (N+1, M) and (N, M+1) for us to observe the PAT process. We set our microwave at several GHz with power of -5dbm and bring it to bear on the right plunger gate 4. Fig.2 (a-c) shows the DQD phase diagram with a PAT process, where many stripes paralleling to the (N+1, M) to (N, M+1) electron charge transition line [indicated by black dashed line in Fig. 2 (a) and Fig. 2 (b)], are induced by the applying microwave

The order of a PAT process can be counted directly by the number of stripes on one side of the electron transition line. Using microwave with 12GHz frequency, three lines appear on each side of the electron transition line. This is consistent with previous experimental observations of the PAT[13,14] process. When the frequency is increased to 20GHz, the number changes to 5 as shown in Fig. 2 (b). Upon increasing frequency, in addition to increasing photon energy, the lines correspond to electron jumping to the left/right reservoir are also broadened and the background current noise is amplified as well. For the case where the microwave frequency is 30GHz, we did not observed a clear picture but just an obscure one as displayed in Fig. 2 (c).

To confirm that this phenomenon is a PAT process, we have measured the microwave frequency dependence of the resonance. Theory offers a simple and clear relation between frequency and splitting distance between two neighboring peaks[17]:

$$\alpha\varepsilon = \sqrt{(hf)^2 - (2t)^2} \tag{1}$$

where α is level arm of gate, ε is detuning and 2t is the interdot coupling energy. We fit the data using this equation in Fig. 2(d). The curve is nearly linear at large frequency, which proves that what we observe is a PAT process. The fitting parameter for the level arm is about 0.04 and for the interdot coupling is about 8GHz.

Upon raising the microwave power, the probability for an electron to absorb more than one photon increases, and higher order PAT lines should appear. We cannot finish a long-time sweeping when the microwave power is very high since it heats the He3 pot and the system would be warmed up in less than one hour. Consequently, a compromised method is employed to study the PAT process in the presence of high microwave power. By sweeping along a line crossing the charge transition line perpendicularly, we can still extract information about the PAT process such as height, splitting distance and number of peaks from the current read by the QPC. Fixing the line along which the sweeping occurs, we can study how the process evolves by comparing differences of the measured traces and deduce the changes due to different experimental conditions, such as microwave power.

In Fig. 3, the valley on the curve corresponds to the situation when the power is off. It separates the charge state (N+1, M) and (N, M+1) ($\varepsilon = 0$, indicated by the blue dashed line). As microwave power increases, focusing on the 15 GHz frequency, more peaks (up to 14 photon at 10dbm) appear on the right side of the blue dashed line. Every point on the curve is ten-times averaged in one scan. When the power is 5dbm, the measured electron temperature is about 2K. As a result, we estimate that the electron temperature in this sample is 3-4K at 10dbm in view of the fact that 10dbm is 3 times larger than 5dbm.

LZS interference pattern with detuning and microwave power appears in a strongly and periodically driven two-level system. It has been observed in many systems such as Cooper-pair box and semiconductor quantum dot[9,19,20]. The evolution of traces shown in Fig. 3, where we increase the power of the sine-shaped continuous wave (cw), could be understood as LZS interference rather than the usual PAT process. In Fig. 3, successive traces with different power are offset by 20pA so they can be drawn clearly in a single figure. Parallel short lines define the height h of a peak, which changes with microwave power. The height of the 1γ peak (indicated by the red dashed line) oscillates with microwave power from -4dbm to 4dbm as shown in the inset of Fig. 3. In the vicinity of -4dbm, 0dbm and 4dbm, the height of the 1γ peak almost falls down to zero. The 2γ-6γ peaks exhibit the same behavior (not shown in the Fig. 3). The 7γ peaks and those which correspond to more photon absorption rise up obviously when the power is increased to 10dbm. As discussed above, we are not able to finish long time sweepings and provide a complete picture because of the cryogenic platform we use. However, the curves at different powers have three signatures which suggest that this process is indeed the LZS interference: firstly, heights of the peaks oscillate with power and nearly decrease to zero for some special values of power; secondly, more peaks appear at large detuning when microwave power increases; thirdly, the splitting distance between peaks is uniform. These signatures strongly favor the interpretation in terms of the LZS interference even when the electron temperature is as high as several Kelvins.

An important parameter in the study of quantum computation is the $T_1$ time, which quantifies how long it takes for an electron on the upper level to transit to the lower level. In the semiconductor heterostructure DQD system we use, the charge state relaxation time could be extracted from the resonance of the two-level system. We are interested in whether this parameter in our two-level system remains unchanged or not when the system is driven out of thermal equilibrium and has high electron temperature.

To measure $T_1$, we should first modulate the microwave signal. We use a mixer (type: Marki M80420MS 1120) and a function generator (type: Agilent E8257D) to realize a chopper which cuts microwave amplitude to zero in half of a repeat period time $\tau$ (shown in Fig. 4). We define $M(\tau)$ as the ratio of the height of a PAT peak (in unit of the charge occupation number between 0 and 1) at different $\tau$ to the height at $\tau=0$. For $\tau \gg T_1$, the time is long enough for the electron to jump back to the low-energy level so the QPC gives an average value $M = 0.5$. When $\tau$ gets closer to zero, the electron is very likely to be excited once again and has no time to move to the low-energy level so we can set $M(\tau = 0) = 1$ in this case. As $\tau$ increasing, heights of the PAT peaks decrease and we can measure the ratio $M(\tau)$ of the heights of the peaks at different $\tau$. However, we find that, after measuring the output microwave using an oscillation, the microwave shape cannot be kept as sine function when $\tau$ is shorter than 3ns (300MHz). Consequently, we replace the condition $M(\tau = 0) = 1$ with $M(\tau = 3ns) = 1$. Here we set the chopped microwave power to 12dbm and the measured electron temperature is above 1K. There is an exponential decay relation between $M$ and $\tau$ in the presence of modulated cw microwave given by the following equation[14,18]:

$$M(\tau) = \frac{1}{2} + \frac{T_1(1-e^{-\tau/2T_1})}{\tau} \qquad (2)$$

In Fig. 4, we get $T_1$=8ns when fitting the data using this equation, which agrees with the measurements reported by others[14,16], although the electron temperature (above 1K) are far above the lattice temperature (which is kept at about 250mK). We speculate that the relaxation rate due to electron-phonon interaction mainly depends on the lattice temperature instead of the electron temperature.

In conclusion, we report a PAT process of a charge qubit far from thermal equilibrium. PAT peaks as high as 14th order are observed and LZS signatures are demonstrated even when the electron temperature is more than 2K. We get ~8ns for the charge state relaxation time $T_1$, which is consistent with previously reported values in thermal non-equilibrium cases.

**Acknowledgements:** This work was supported by the National Fundamental Research Program (Grant No. 2011CBA00200), NNSF (Grant Nos. 11222438, 10934006, 11274294, 11074243, 11174267 and 91121014), and CAS.

**References:**

[1] M. Ciorga, A.S.Sachrajda, P. Hawrylak, C.Gould, P. Zawadzki, S. Jullian, Y. Feng and Z. Wasilewski, *Phys. Rev. B* **61**, 16315 (2000)

[2] L.P.Kouwenhoven, D.G. Austing, and S.Tarucha, *Rep. Prog. Phys.* **64**, 701(2001)

[3] W.G.van der Wiel, S.De Franceschi, J.M. Elzerman, T.Fujisawa, S.Tarucha, L. P. Kouwenhoven, *Rev. Mod. Phys.* **75**, 1 (2002).

[4] Toshimasa Fujisawa, Toshiaki Hayashi1 and Satoshi Sasaki, *Rep. Prog. Phys.* **69**, 759 (2006).

[5] R.Hanson, L.P.Kouwenhoven, J.R.Petta, S.Tarucha, and L.M.K. Vandersypen, *Rev. Mod. Phys.* **79**, 1217(2007).

[6] L. P. Kouwenhoven, S. Jauhar, J.Orenstein, and P.L.McEuen, *Phys. Rev. Lett.* **73**,3443(1994)

[7] T.H.Oosterkamp, L.P.Kouwenhoven, A.E.A.Koolen, N.C van der Vaart, and C.J.P.M. Harmans, *Phys. Rev. Lett.* **78**,1536(1997)

[8] T.H.Oosterkamp, T.Fujisawa, W. G. van der Wiel, K. Ishibashi, R.V.Hijman1, S.Tarucha and L.P.Kouwenhoven1, *Nature* **395**, 873(1998)

[9] J.Stehlik, Y.Dovzhenko, J.R.Petta, J.R.Johansson, F.Nori, H.Lu, and A. C.Gossard, *Phys. Rev. B* **86**,121303(2012)

[10] Helium3 refrigeration (type: Heliox VL) manufactured by Oxford Instrument

[11] E.B.Foxman, U.Meirav, P.L.McEuen, M.A. Kastner, O.Klein, P.A.Belk, and D.M.Abusch, *Phys. Rev. B*, **50** 14193(1994).

[12] M.Ciorga, A.S.Sachrajda, P.Hawrylak, C.Gould, P.Zawadzki, S.Jullian, Y. Feng, and Z.Wasilewski1, *Phys. Rev. B* **61**,16315(2000)

[13] J.M.Elzerman, R.Hanson, J.S.Greidanus, L.H.Willems van Beveren, S.De Franceschi, L.M.K.Vandersypen,S.Tarucha, and L.P.Kouwenhoven , *Phys. Rev. B* **67**,161308(2003)

[14] J.R.Petta, A.C.Johnson, C.M.Marcus, M.P.Hanson, and A.C. Gossard, *Phys. Rev. Lett.* **93**,186802(2004)

[15] E. Dupont-Ferrier, B.Roche, B.Voisin, X.Jehl, R.Wacquez, M.Vinet, M.Sanquer, and S.De Franceschi, *Phys. Rev. Lett.* **110**,136802(2013)

[16] O.E.Dial, M.D.Shulman, S.P.Harvey, H.Bluhm, V.Umansky, and A.Yacoby, *Phys. Rev. Lett.* **110**,146804(2013)

[17] W.G. van der Wiel, T. Fujisawa, T.H. Oosterkamp, L.P. Kouwenhoven, *Physica B* **272**,31(1999)

[18]K.W. Lehnert, K. Bladh, L. F. Spietz, D. Gunnarsson, D.I. Schuster, P. Delsing, and R.J. Schoelkopf, *Phys. Rev. Lett.* **90**, 027002(2003)

[19]William D. Oliver, Yang Yu, Janice C. Lee, Karl K. Berggren, Leonid S. Levitov, Terry P. Orlando, *Science* **310**, 1653(2005)

[20] C.M. Wilson, T. Duty, F. Persson, M. Sandberg, G. Johansson, and P. Delsing, *Phys. Rev. Lett.* **98**, 257003(2007)

**Figure captions**:

**Figure1.** (a) The Ti/Au gates 1 to 6 define a typical double quantum dot pattern and gate 7 is designed to modulate the QPC current. Gate 2 and 4 are called as plunge gates. QPC current flows from the source 8 (a point on the top presents that terminal will be connected to the ohmic contact) to drain 9. (b) The figure shows a honey-comb area with four different charge states (N, M), (N+1, M), (N, M+1) and (N+1, M+1) detected by QPC measurement.

**Figure2.** (a-c) are measured under power of -5dbm. In (a) and (b), the black dashed line indicates the position of $\varepsilon =0$ (0γline). (a) and (b) shows QPC differential current dI/dV honey-comb diagram with different microwave frequency. 3 and 5 paralleled lines appear on each side with the frequency of 12GHz and 20GHz. In (c), microwave photons of 30GHz will obviously broaden the lines tunneling to the reservoir and amplify the background current noise so that whole picture becomes obscure. In (d), a fitting curve between the microwave frequency and splitting distance proves the phenomenon we observed as a PAT process.

**Figure3.** Differential current dI/dV curves crossing the interdot tunneling line with different microwave power at frequency of 15GHz are shown and every point on the curves is ten-times averaged in one scan. Blue dashed line illustrates the position of zero detuning. Successive traces (power at off, 0bm, 6dbm and 10dbm from up to down) are offset by 20pA. Double paralleled short lines define h as the height of peak. Inset shows that 1γpeak (indicated by red dashed line) height performs an oscillation between 3pA and 12pA with microwave power and reach the minimum in the vicinity of -4dbm, 0dbm and 4dbm. Peaks 2γ-6γperform the same behavior as well. Height of peaks 7γ and more photons obviously rises up by increasing the power to 10dbm.

**Figure4**. Ratio of resonance amplitude, compare to $\tau = 3ns$ is an exponential function of period τ at frequency of 15GHz and power of 12dbm. The fitting curve provides the $T_1 = 8ns$.

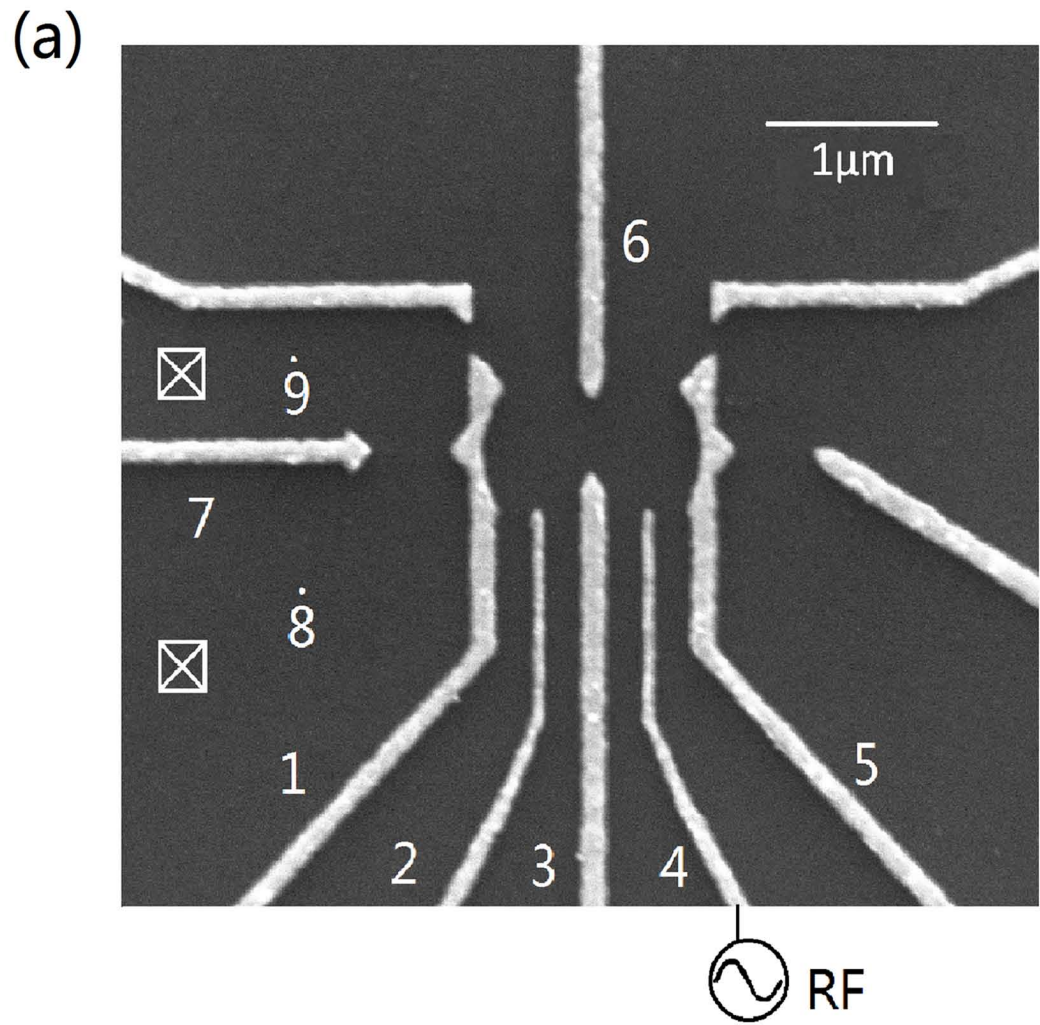 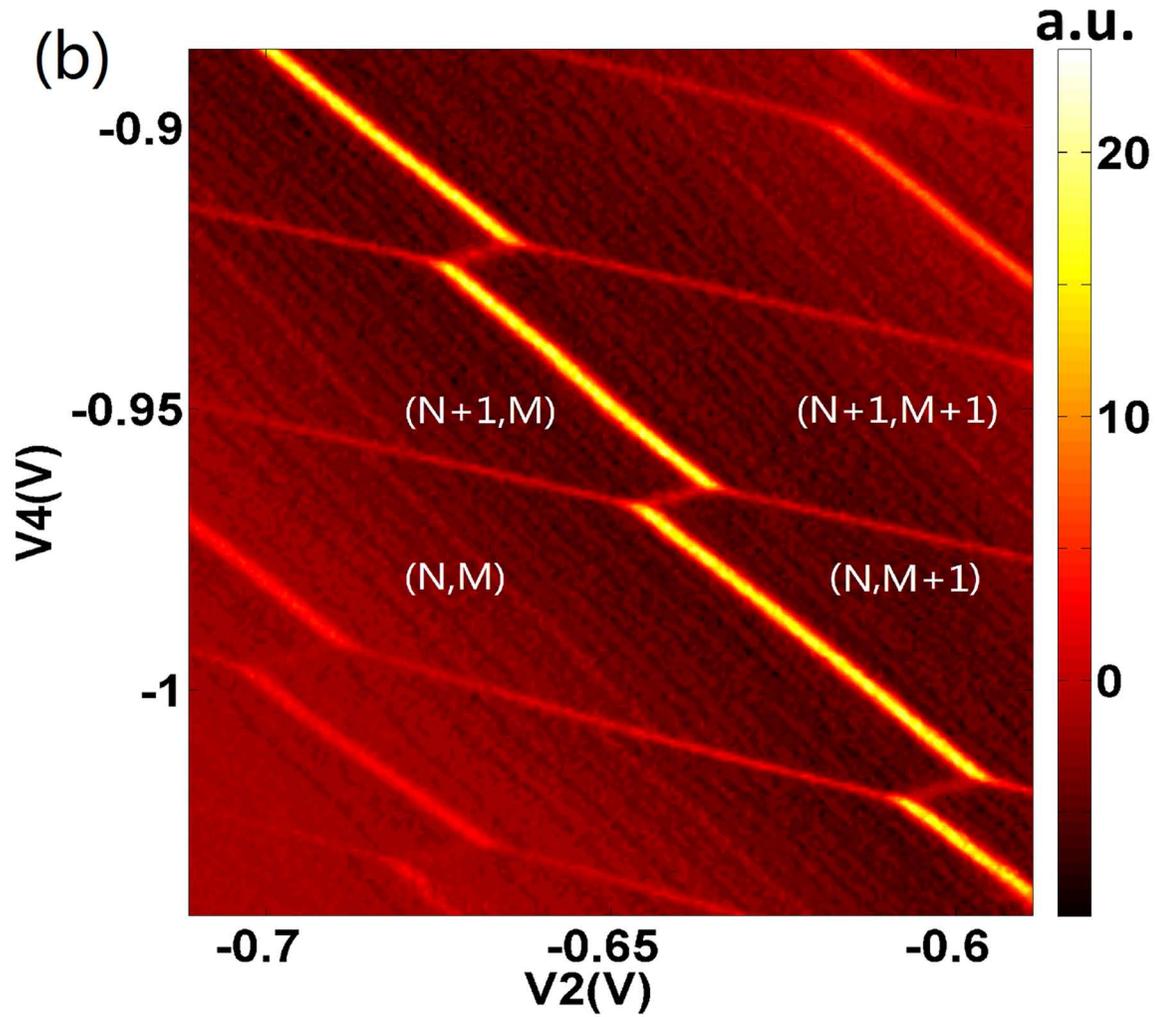

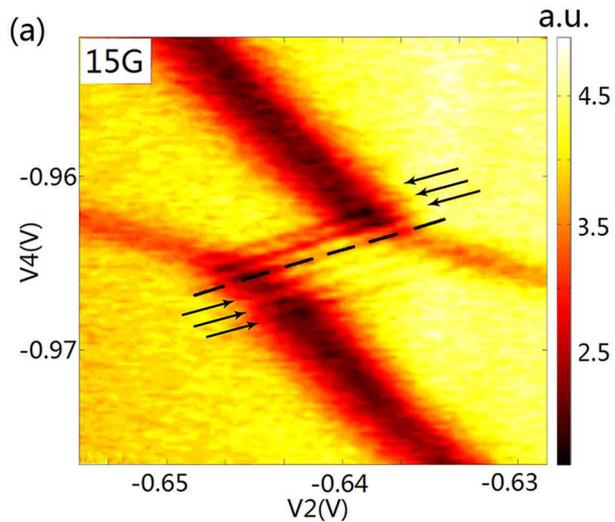 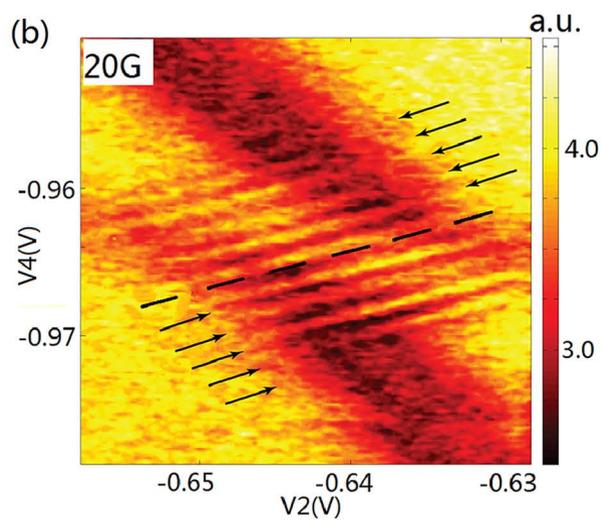
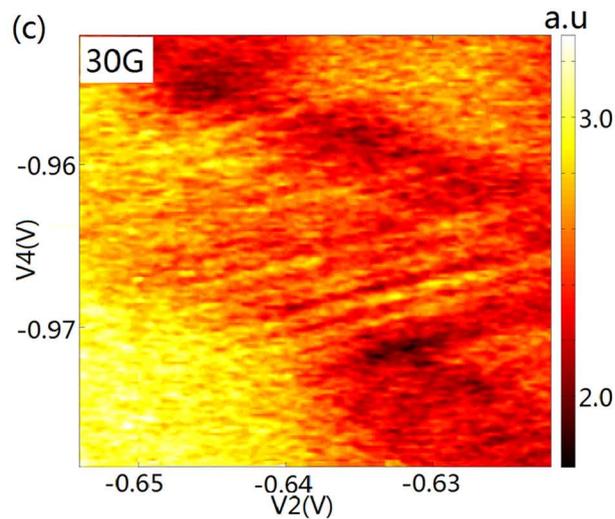 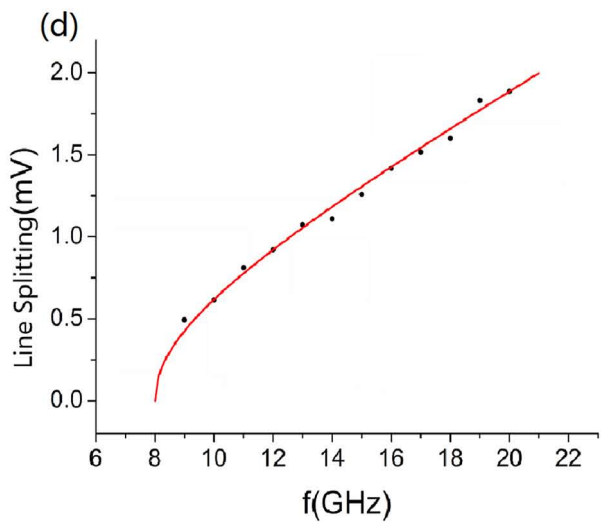

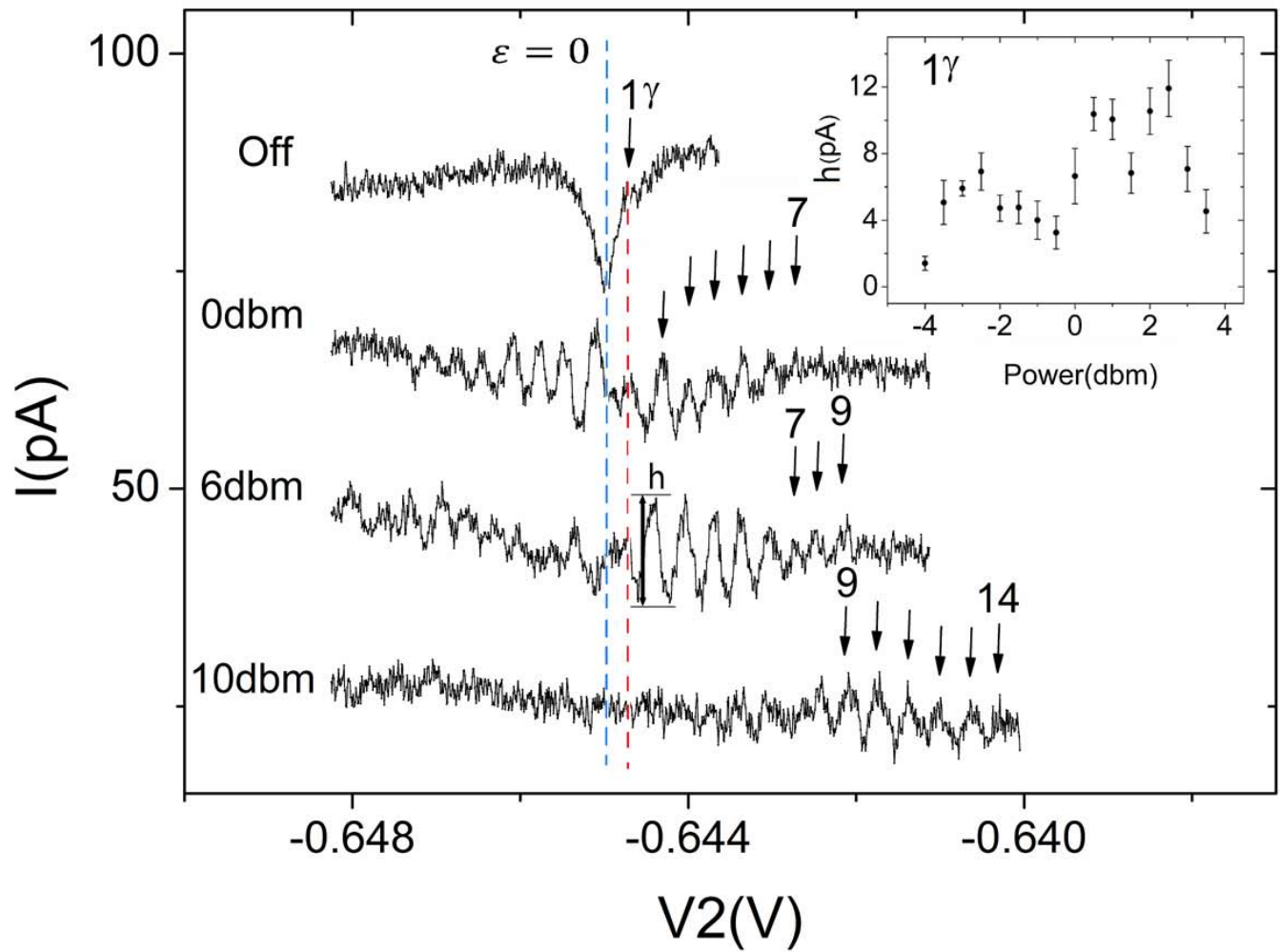

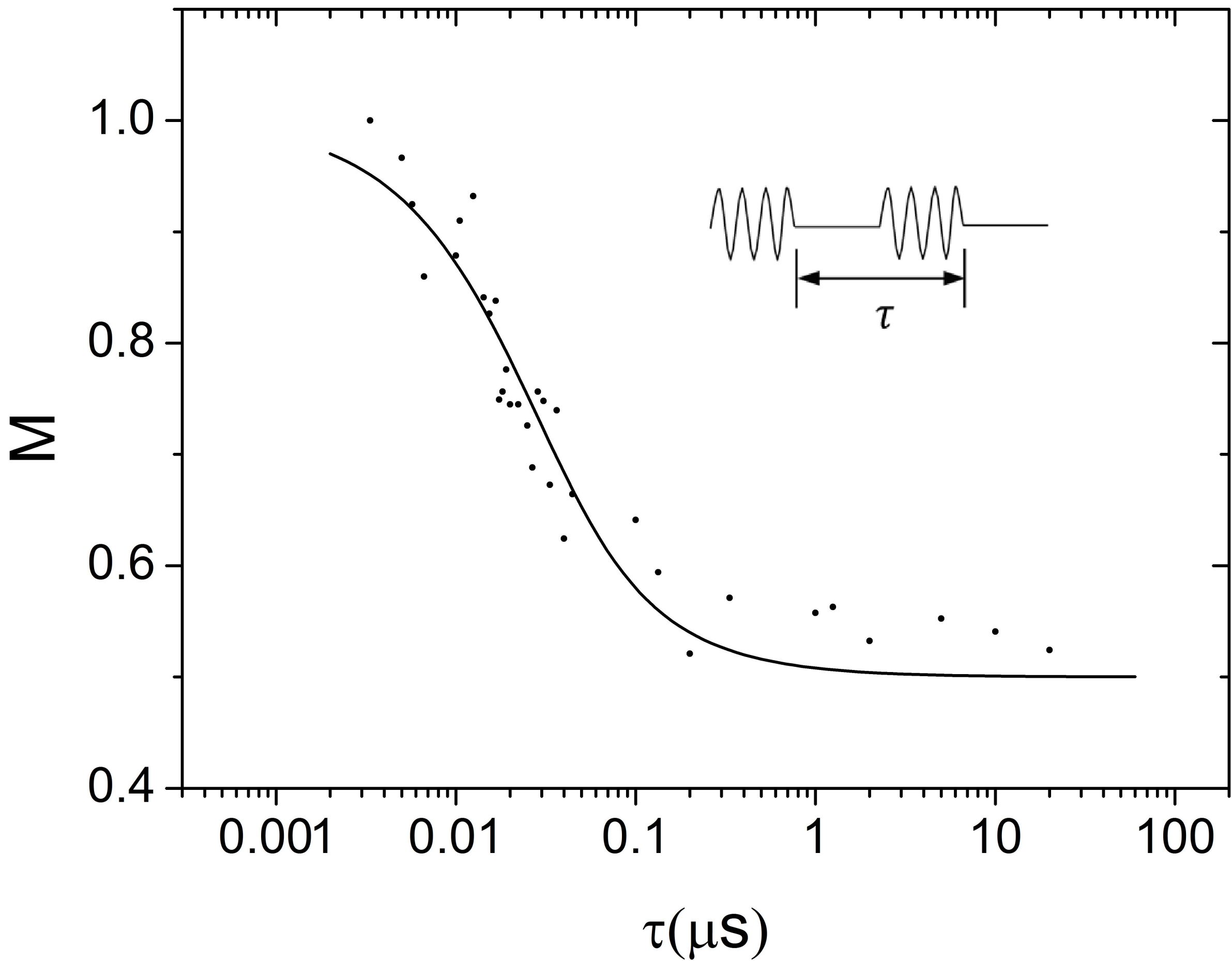